\documentclass[10pt,journal,compsoc]{IEEEtran}
\ifCLASSOPTIONcompsoc
  \usepackage[nocompress]{cite}
\else
  \usepackage{cite}
\fi
\ifCLASSINFOpdf
\else
\fi
\usepackage[most]{tcolorbox}
\newcommand{\R}{\mathbb{R}}

\newcommand{\x}{\mathbf{x}}

\newcommand{\wv}{\mathbf{w}}

\usepackage{nccbbb}
\usepackage{graphicx,subfigure}
\usepackage{mdwlist}
\usepackage{array}
\usepackage{multirow}
\usepackage{float}
\usepackage{listings}
\usepackage{enumitem}
\usepackage{amsmath,amssymb,amsfonts}
\usepackage{textcomp}
\usepackage{subfigure}
\usepackage{underscore}
\usepackage{algorithm}
\usepackage{algpseudocode}
\usepackage{graphics}
\usepackage{epsfig}

\expandafter\let\csname algorithm*\endcsname\relax
\expandafter\let\csname endalgorithm*\endcsname\relax

\usepackage[ruled,vlined,linesnumbered]{algorithm2e}
\usepackage[most]{tcolorbox}
\begin{document}
\title{SCADS: A Scalable Approach Using Spark in Cloud for Host-based Intrusion Detection System with System Calls}

\author{Ming~Liu,
        Zhi~Xue,
        Xiangjian~He,~
        and~Jinjun~Chen
\IEEEcompsocitemizethanks{
\IEEEcompsocthanksitem M. Liu is with the School of Electronics, Information and Electrical Engineering, Shanghai Jiao Tong University, Shanghai, China.
\IEEEcompsocthanksitem Z. Xue is with the School of Electronics, Information and Electrical Engineering, Shanghai Jiao Tong University, Shanghai, China.
\IEEEcompsocthanksitem X. He is with the Faculty of Engineering and Information Technology, University of Technology Sydney, NSW, Australia.
\IEEEcompsocthanksitem J. Chen is with the Faculty of Science, Engineering and Technology, Swinburne University of Technology, VIC, Australia}

\thanks{}}

\markboth{}
{Shell \MakeLowercase{\textit{et al.}}: SCADS: A Scalable Approach Using Spark in Cloud for Host-based Intrusion Detection System with System Calls}

\IEEEtitleabstractindextext{%
\begin{abstract}
Following the current big data trend, the scale of real-time system call traces generated by Linux applications in a contemporary data center may increase excessively. Due to the deficiency of scalability, it is challenging for traditional host-based intrusion detection systems deployed on every single host to collect, maintain, and manipulate those large-scale accumulated system call traces. It is inflexible to build data mining models on one physical host that has static computing capability and limited storage capacity. To address this issue, we propose SCADS, a corresponding solution using Apache Spark in the Google cloud environment. A set of Spark algorithms are developed to achieve the computational scalability. The experiment results demonstrate that the efficiency of intrusion detection can be enhanced, which indicates that the proposed method can apply to the design of next-generation host-based intrusion detection systems with system calls.
\end{abstract}

\begin{IEEEkeywords}
Apache Spark, cloud computing, big data, intrusion detection, system call
\end{IEEEkeywords}}

\maketitle
\IEEEdisplaynontitleabstractindextext
\IEEEpeerreviewmaketitle

\raggedbottom
\IEEEraisesectionheading{\section{Introduction}\label{sec:introduction}}
\IEEEPARstart{H}{ost-based} intrusion detection system (HIDS) is renowned for the fine-grained analysis and the capability of discovering internal malicious behaviors. HIDS monitors logs from operating systems, whereas the network-based intrusion detection system (NIDS) focuses on the data flow of network traffic. HIDS with system calls (or system call-based HIDS) investigates gathered system call traces from Unix-like operating systems. This variety of HIDS initiated decades ago, and it has shown the ability to identify intrusions in Unix-like hosts. Conventionally, system call-based HIDS is often deployed on a physical machine to analyze malicious behaviors. It is challenging to manage and process the increasing quantity of system call traces originating from Unix-like operating systems of numerous virtual machines in a data center. Massive system call traces are a form of big data, which may be complex to be efficiently analyzed by standalone data mining methods. Training machine learning classifiers with big data on one individual physical machine that possesses fixed computational resources may require excessive time. The scale of intermediate datasets may not be tolerable for the internal memory of a PC. If those datasets are saved on hard drives, the real-time analysis may be postponed because of the I/O speed. In this case, it is demanded to propose innovative and cost-effective HIDS schemes for some particular Linux applications.

Recently, cloud computing presents extensive computational capability and massive storage capacity that can facilitate security specialists to implement data-intensive projects with manageable expenditure. Users can focus on their works using a group of flexible IT services, with less concern about the purchase and maintenance of physical devices. Therefore, various security enterprises have moved their projects to the cloud. 
Moreover, a set of modern frameworks such as Apache Hadoop and Apache Spark are specifically developed for stable and scalable processing of big data. These frameworks enable the processing and storage of massive datasets among clusters that have the "master-workers" structure. Clusters can be created with multiple common computers, which provide local computation and storage capability. With these big data processing frameworks, computational resources in clusters can be scheduled, hardware failures can be handled, and functional user interfaces can be provided. Therefore, combining those big data processing frameworks and the capability of cloud computing can provide an opportunity to improve the detection efficiency of traditional system call-based HIDS.

To the best of our knowledge, there are limited research works concerning applying big data tools such as Apache Spark to system call-based HIDS. Motivated by this issue, in this article, we contribute to the community of HIDS by proposing a scalable HIDS approach using Spark in the Google cloud, endeavoring to improve the detection efficiency and the scalability for a new-generation system call-based HIDS. 

The rest of this article is organized as follows. In Section 2, we provide related works about host-based intrusion detection system with system calls. In Section 3, we introduce the design of SCADS, the scalable approach using Spark in the Google cloud for HIDS with system calls. In Section 4, we demonstrate the experiment results. In Section 5, the conclusion is provided, and potential future works are discussed.

\section{Related works}
\subsection{Host-based intrusion detection system with system calls}
System call-based HIDS was initially proposed as the ``sequence time-delay embedding (STIDE)" method \cite{forrest1996sense}\cite{hofmeyr1998intrusion}\cite{warrender1999detecting}\cite{forrest2008evolution}\cite{9342968}. In STIDE, system call arguments are ignored. Short system call sequences from normal processes are used to construct databases of normal behaviors; system call sequences for testing are compared with the normal databases to identify anomalies. Since then, researchers have been devoting to improving the detection accuracy of system call-based HIDS. When it comes to rule learning-based HIDS methods \cite{lee1997learning}\cite{lee1998data}\cite{mahoney2003learning}\cite{jiang2007multiresolution}\cite{tandon2008machine}\cite{ye2010intrusion}, short system call sequences are extracted and labeled to train a set of rules, and testing sequences that differ from those predefined rules are considered as anomalies \cite{guan2009fast}. It is also notable that the Hidden Markov Model (HMM) is broadly adopted in system call-based HIDS, where the evaluation, decoding, and learning problems of HMM are applicable \cite{warrender1999detecting}\cite{yeung2003host}\cite{qiao2002anomaly}\cite{hu2010host}\cite{haider2017detecting}. However, the main drawback of the HMM-based methods is the high time-complexity of training. Other implementations include k-means \cite{xie2013evaluating}\cite{xie2014evaluatinga}, k-nearest neighbor \cite{liao2002using}\cite{Yuxin2011}, support vector machine \cite{xie2014evaluatingb}\cite{Watson2016}\cite{khreich2017combining}, Bayesian networks \cite{Mutz2006}, decision trees \cite{Arshad2013b}, neural networks \cite{Abdel-Azim}\cite{das2016semantics}\cite{Lichodzijewski}\cite{Ahmed}\cite{chen2015self}\cite{chen2017anrad}\cite{Ghoshb}, and Bloom filter \cite{Wang}, etc. 

Besides the detection accuracy, the detection efficiency is another important factor. Low detection efficiency may result in the delay of real-time intrusion detection from a large amount of system call traces. Therefore, researchers also have been attempting to reduce the processing cost and computational complexity. For instance, Hu et al. provided an approach which eliminates near-duplicate sequences of system calls to accelerate the training process of HMM \cite{hu2009simple}. A novel ``nested-arc hidden semi-Markov model (NAHSMM)" was developed by Haider et al. to improve the training efficiency of HMM \cite{haider2017detecting}. The ``Kernel State Modeling" approach was proposed by Murtaza et al. to transfer system call traces to kernel module traces for the reduction of processing time \cite{murtaza2013host}\cite{murtaza2015trace}. The ``anomaly recognition and detection (AnRAD)" approach proposed by Chen et al. achieves high-speed incremental learning of data streams by implementing ``a real-time self-structuring learning framework" \cite{chen2015self}\cite{chen2017anrad}, etc.

\subsection{Public cloud}
Contemporary public cloud computing services can offer load-balanced IaaS (Infrastructure as a Service) for scalable data analytics. Based on the information provided by Amazon EC2 \cite{walker2008benchmarking} and Google Compute Engine \cite{krishnan2015google}, IaaS cloud computing services can be scaled from single to multiple virtual machines (instances) running in data centers with consistent performance. Users can manage and configure those instances on web interfaces. The pricing of those public cloud computing services is flexible. According to the scale of their projects, users can choose and pay for the sizes of processors, the volume of storage, and the time of computation. Moreover, renowned enterprises such as Amazon and Google may provide more reliable and less vulnerable cloud computing services, considering intruders may attack the intrusion detection software based on the vulnerabilities. Therefore, system call-based HIDS can be deployed in the public cloud. General acceleration methods used in the community for system call-based HIDS can also be integrated into public cloud platforms.

\subsection{Apache Spark}
Contemporary big data processing frameworks (or big data tools) refer to a group of open-source software that can execute on clusters built with common hosts for distributed, scalable, and reliable processing of large-scale datasets. These frameworks provide functionalities such as distributed storage and processing of big data, and these frameworks usually have user-friendly APIs and web interfaces. These frameworks can deliver high-availability services, and frequent hardware failures can be automatically handled and restored. Distributed algorithms can be passed to worker nodes to process the local data. Therefore, with big data tools deployed on clusters, datasets can be efficiently processed.
\begin{figure}[h]
	\centering
	\includegraphics[width=0.45\textwidth]{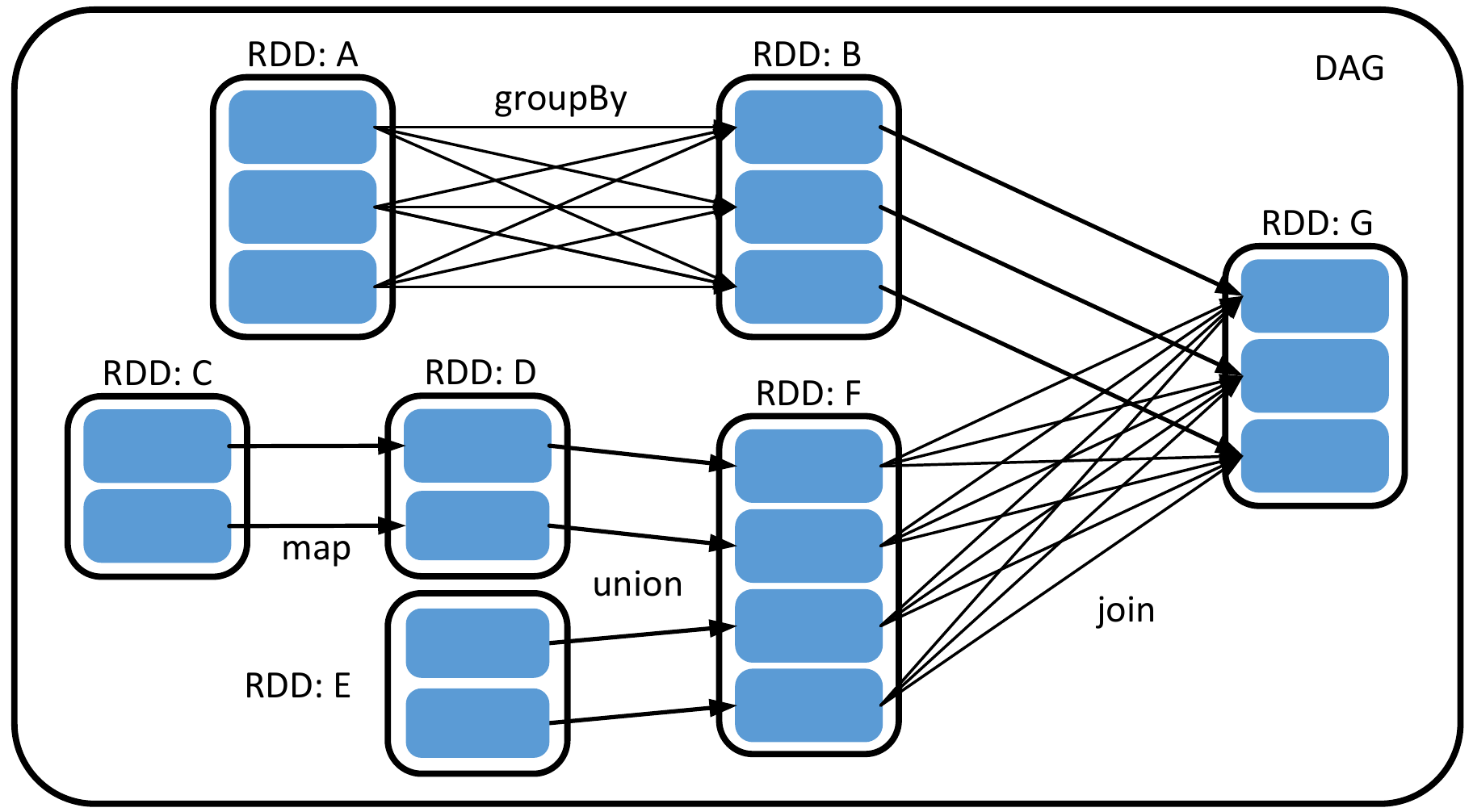}
	\caption{An example of RDD and DAG. Unshaded rounded rectangles are RDDs, shaded rounded rectangles are partitions. Modified from Zaharia's article \cite{zaharia2012resilient}.}
	\label{RDDnDAG}
\end{figure}
\begin{figure}[h]
	\centering
	\includegraphics[width=0.4\textwidth]{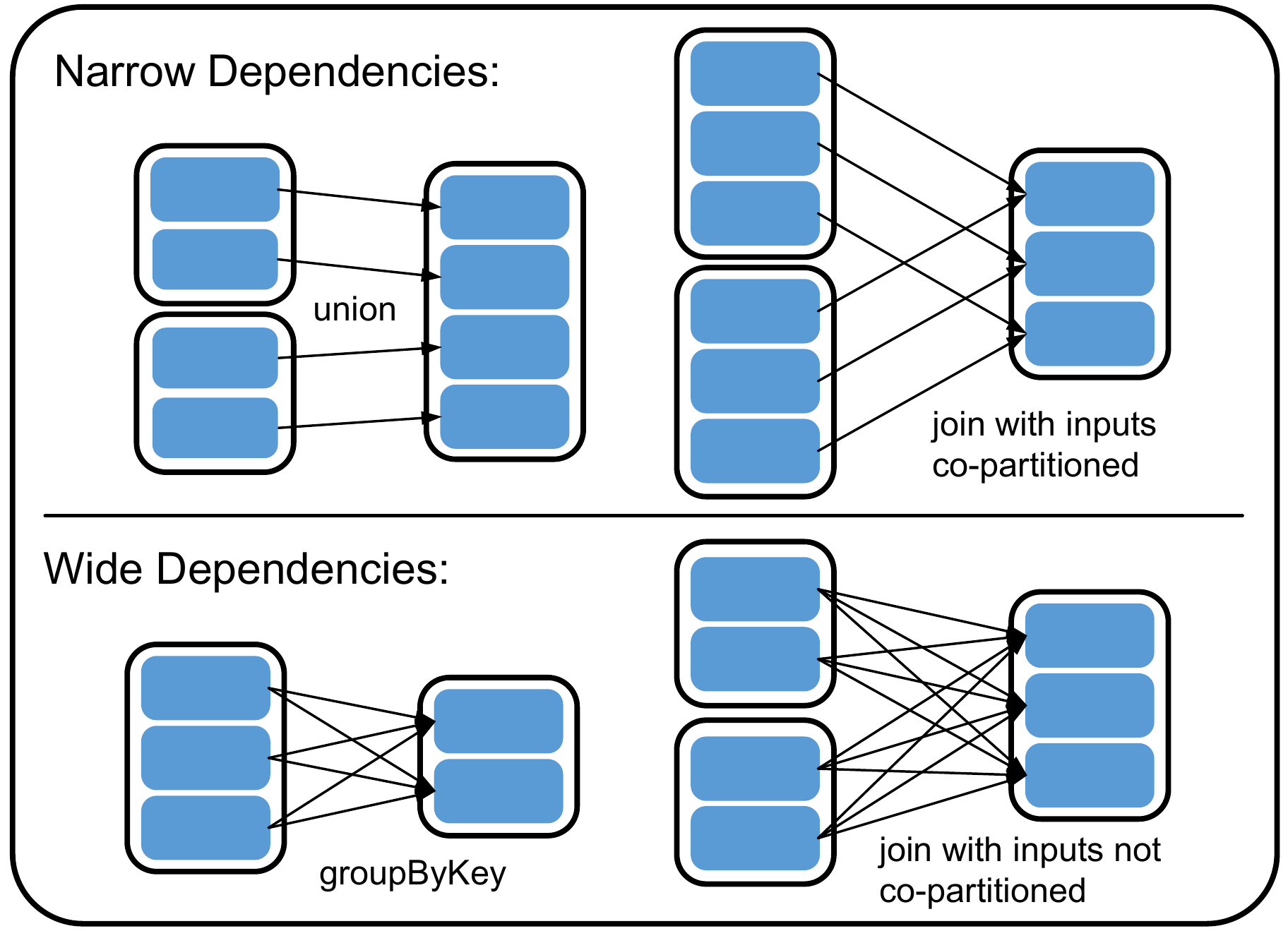}
	\caption{Examples of narrow dependencies and wide dependencies. Modified from Zaharia's article \cite{zaharia2012resilient}.}\label{NarrowWide}
\end{figure}
\begin{figure*}[t]
	\centering
	\includegraphics[width=\textwidth]{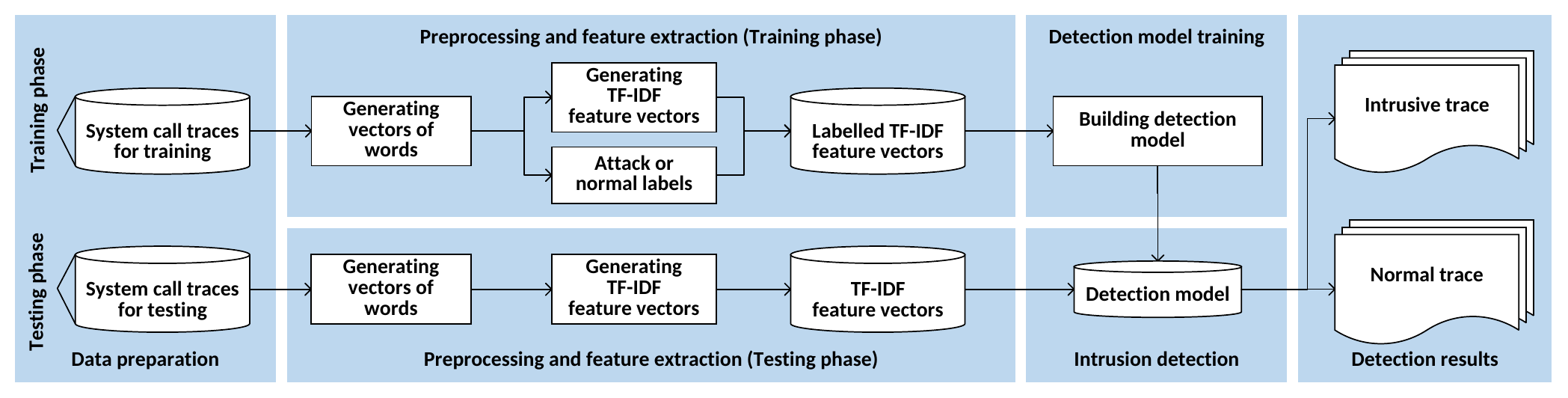}
	\caption{The training and detection processes of SCADS.}\label{SCADS}
\end{figure*}

Apache Spark is an open-source in-memory cluster computing framework \cite{shanahan2015large}. Spark can exchange data with various distributed storage systems such as the Hadoop Distributed File System (HDFS), which splits files into redundant blocks that are distributed among worker nodes. Spark facilitates the implementation of iterative algorithms (such as the training of machine learning models). The intermediate datasets can be cached in the distributed memory. The distributed and fault-tolerant computing paradigm of Spark is based on the Resilient Distributed Dataset (RDD) and the relevant directed acyclic graph (DAG) (depicted in figure \ref{RDDnDAG}). 

An RDD consists of multiple partitions. The dependencies between RDDs can be classified into two types, namely, narrow dependencies and wide dependencies. Narrow dependencies allow pipelined executions in one cluster node with no shuffling, whereas wide dependencies require data shuffling across nodes \cite{zaharia2012resilient}. Some examples of narrow dependencies and wide dependencies are depicted in figure \ref{NarrowWide}. As the process of data shuffling in wide dependencies may require additional processing time, unnecessary data shuffling should be avoided.

\section{Scalable Approach Using Spark in Cloud for system call-based HIDS}
The intrusion detection procedure for SCADS is depicted in figure \ref{SCADS}. There are two main phases in this procedure, i.e., the training phase and the testing phase. There are several steps for each phase.
\begin{enumerate}[label=(\arabic*),leftmargin=0em,itemindent=1.5em,labelwidth=\itemindent,labelsep=1mm]
	\item \textbf{The training phase}. Firstly, the ADFA-LD dataset is prepared and stored in the cloud storage system. For the preprocessing and feature extraction of the training phase, vectors of words are generated from raw system call traces with the multiple-length n-gram method. The single-length n-gram method is also applied for comparison. TF-IDF feature vectors are generated from the vectors of words and labeled as normal or attack. Two kinds of classifiers are trained in the experiments. One is the logistic regression with limited-memory BFGS (LR-LBFGS), the other is the linear support vector machine (SVM).
    \item \textbf{The testing phase}. For the preprocessing and feature extraction of the testing phase, vectors of words are generated from raw system call traces with the multiple-length n-gram method. The single-length n-gram method is also applied for comparison. For the generation of TF-IDF feature vectors, it is assumed that in the real-world scenario of intrusion detection, real-time system call traces are gathered by the IDS trace by trace. Therefore, the IDF model of the testing dataset cannot be obtained at one-time. In this case, for each testing trace, the TF vector is generated, then the TF-IDF feature vector is generated using the IDF model of the training dataset. Then the TF-IDF feature vector is passed to the detection models (LR-LBFGS and SVM) for the prediction of either intrusive or normal trace. Finally, the value of AUC is returned to evaluate the detection accuracy. TF feature vectors are used in this phase for comparison. The detailed steps are elaborated in the following subsections. 
\end{enumerate}

\subsection{Symbols}
Symbols used for the description of the proposed approach are listed and described in table \ref{symbols}.

\begin{table}[h]
	\centering
	\caption{Symbols and descriptions.}
	\label{symbols}
	\begin{tabular}{|p{34pt}|p{185pt}|}
		\hline
		Symbols& 
		Descriptions \\
		\hline
		$\bbbh$ & 
		The complete dataset of system call traces $\mathcal{T}$\\
		$\mathcal{T}$& 
		A trace of system calls $S$\\
		$S$& 
		A system call\\
		$\bbbd$ & 
		The complete corpus of documents $\mathcal{D}$\\
		$\mathcal{D}$ & 
		The generated document that contains the extracted words $W$ of system calls $S$\\
		$W$ & 
		An extracted word of system calls $S$\\
		\hline
	\end{tabular}
\end{table}

\subsection{Preprocessing and feature extraction}
\subsubsection{RDD repartitioning}
To achieve the detection efficiency and scalability for a HIDS using Spark, the design of the intrusion detection algorithms should conform to the internal mechanism of Spark. To achieve distributed processing, the raw system call traces should be firstly distributed evenly among worker nodes of the Spark cluster, and then be processed by appropriate algorithms in parallel. In the proposed approach, after eliminating redundant traces, the raw system call traces for training and testing are treated as Spark RDDs, each RDD has several partitions. When the system call traces are firstly loaded from the storage system and cached as RDDs, they may only have a few partitions and are not distributed evenly among worker nodes of the cluster. Therefore, we have designed an RDD repartitioning method using the Spark API to transform those RDDs of raw system call traces into new RDDs with more partitions; the new partitions are distributed evenly to the worker nodes.
\begin{algorithm}
	\caption{Generate vectors of words from an RDD of raw system call traces}
	\label{alg2}
	\KwIn{An RDD of system call traces $traces$}
	\KwOut{An RDD of vectors of words $vects$}
	\DontPrintSemicolon
	\SetAlgoLined
	\SetKwProg{Def}{def}{$=$}{end}
	\Def{$\mathrm{VectsGen}(traces: \mathrm{RDD[String]}): \mathrm{RDD[Seq[String]]}$}
	{
		$\mathrm{val}$ $tHashed =$ $traces.\mathrm{map}(x=>(x.\mathrm{hashCode},x))$\\
		$\mathrm{val}$ $rangePtnr =$ $\mathrm{new}$ $\mathrm{RangePartitioner}(traces.\mathrm{count.toInt}, tHashed)$\\
		$\mathrm{val}$ $tPtned =$ $tHashed.\mathrm{partitionBy}(rangePtnr)$\\
		$\mathrm{val}$ $input =$ $tPtned.\mathrm{values.flatMap({\_}.split(''\backslash\backslash s{+}'')).cache}$\\
		$\mathrm{val}$ $output =$ $input.\mathrm{mapPartitions(WordsGen).cache}$\\
		$\mathrm{val}$ $vects =$ $output.\mathrm{mapPartitions}(iter => \mathrm{Iterator}(iter.\mathrm{toArray.toSeq})).\mathrm{coalesce(8,false)}$\\
		\Return{$vects$}\\
	}
\end{algorithm}
\begin{figure*}
	\centering
	\includegraphics[width=0.9\textwidth]{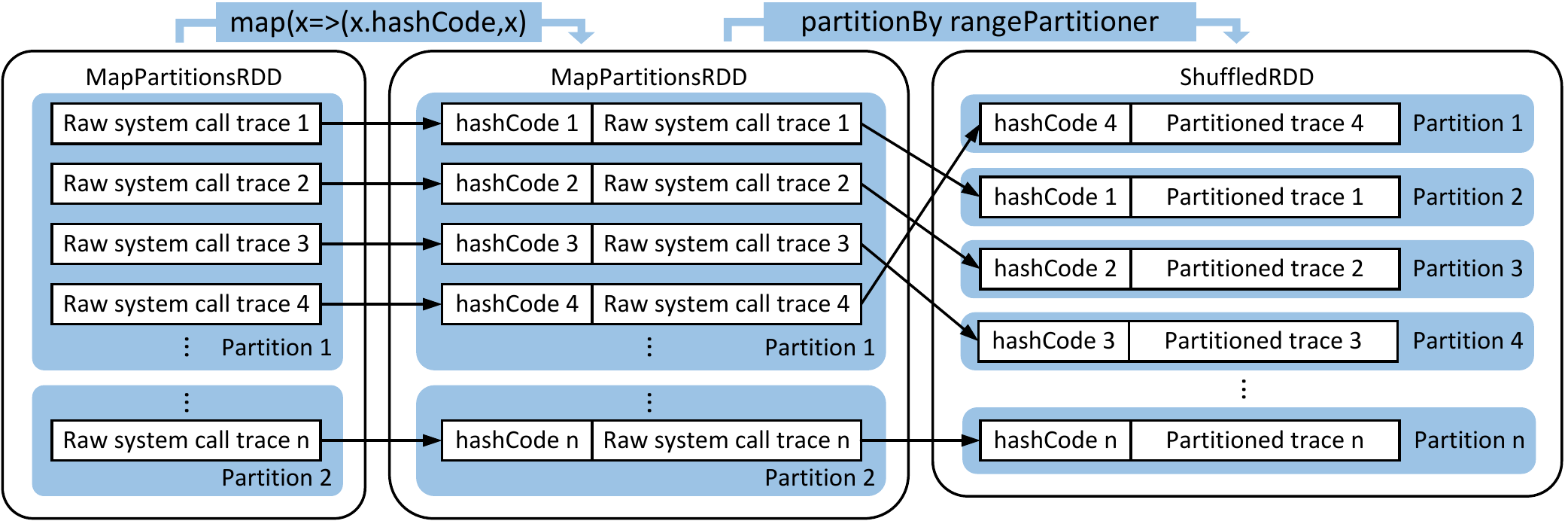}
	\caption{A demonstration of the RDD repartitioning process by \textit{RangePartitioner} for the preprocessing of raw system call traces.\label{repartitioning}}
\end{figure*}

The RDD repartitioning method for system call traces is implemented in algorithm \ref{alg2}, and figure \ref{repartitioning} provides a demonstration of the RDD repartitioning process using the \textit{RangePartitioner} in Spark. The RDD repartitioning method is explained as follows.

\begin{enumerate}[label=(\arabic*),leftmargin=0em,itemindent=1.5em,labelwidth=\itemindent,labelsep=1mm]
	\item For an RDD of system call traces $traces$, each trace is assigned a hashCode value by calling the transformation \textit{map}. Then the RDD $tHashed$ will be repartitioned by the \textit{RangePartitioner}, which can partition traces with hashCode values into roughly equal ranges. 
	\item Then a Spark \textit{RangePartitioner} $rangePtnr$ is constructed with two parameters. The RDD $tHashed$ will be repartitioned by the $rangePtnr$. Since $traces.\mathrm{count.toInt}$ counts the number of system call traces in the dataset, the number of partitions of the new RDD $tPtned$ is roughly equal to the number of system call traces. Thus each partition of $tPtned$ may represent a system call trace in this case.
	\item Then for the repartitioned RDD $tPtned$, after processed by the transformation \textit{flatMap} on the values to split by one or more whitespaces, the transformation \textit{mapPartitions} operates on every partition of the RDD $input$ to extract words of system calls by calling the function \textit{WordsGen}. Then the RDD $output$ is generated.
    \item Then the RDD $output$ is coalesced to eight partitions by using the transformation \textit{coalesce}, which is ``useful for running operations more efficiently after filtering down a large dataset'' \cite{Sparkguide}. The number of coalesced partitions in the proposed approach is selected based on the number of worker nodes. The second parameter of \textit{coalesce} is set to ``false'' in the proposed approach, as it is expected that the coalescence of partitions only occurs within the local worker nodes with no data shuffling, i.e., the coalescence conforms to the narrow dependencies depicted in figure \ref{NarrowWide}. Using the parameter of ``true'' may cause additional shuffling between worker nodes.
\end{enumerate}

\subsubsection{Single-length n-grams}
For system call-based HIDS, an n-gram is referred to a contiguous sequence of $n$ system calls extracted from a system call trace within a time interval \cite{xu2016dynamic}. The sliding window algorithm with window size $n$ can be taken to scan a system call trace to generate n-grams of system calls, and the generated n-grams can be used for training detection models. Using the n-gram method for preprocessing has been widely adopted by HIDS researchers \cite{Wang}\cite{Yuxin2011}.
The term ``single-length n-gram method'' is used to compare with the ``multiple-length n-gram method''.

The process of feature extraction with the single-length n-gram method is described as follows. Let $\bbbh$ = \{$\mathcal{T}_1$,  $\mathcal{T}_2$, ..., $\mathcal{T}_m$\} be the complete dataset of $m$ system call traces $\mathcal{T}$; and let $\mathcal{T}$ = \{$S_1$, $S_2$, ..., $S_k$\} be a trace of $k$ system calls $S$. Then for a trace $\mathcal{T}$, one sliding window of length $n$ is applied to traverse from the beginning to the end of trace $\mathcal{T}$ to extract contiguous system call sequences. The maximum window size $n$ allowed in this case is $k$, which is the length of trace $\mathcal{T}$. The contiguous system call sequences extracted are treated as words $W$ of length $n$. Therefore, if $n$ is equal to 6, for trace $\mathcal{T}$ the generated document $\mathcal{D}$ that contains the extracted words $W$ of system calls is,

\noindent$\mathcal{D}$ = \{$S_1$$S_2$$S_3$$S_4$$S_5$$S_6$, $S_2$$S_3$$S_4$$S_5$$S_6$$S_7$,
$S_3$$S_4$$S_5$$S_6$$S_7$$S_8$,
...,\\ $S_{k-5}$$S_{k-4}$$S_{k-3}$$S_{k-2}$$S_{k-1}$$S_k$\}.

Algorithm \ref{WordsGenSingle} implements the single-length n-gram method using Spark. It takes the sliding window algorithm with one window size $n$. In Algorithm \ref{WordsGenSingle}, $n$ is set to 6. Thus the length of the extracted words $W$ is 6.

\begin{algorithm}
	\caption{Generate words of system calls with the single-length n-gram method from one RDD partition}
	\label{WordsGenSingle}
	\DontPrintSemicolon
	\SetAlgoLined
	\SetKwProg{Def}{def}{$=$}{end}
	\Def{$\mathrm{WordsGen[T:ClassTag]}(iter:\mathrm{Iterator[T]})$}
	{
		$\mathrm{var}\ words = \mathrm{List[String]()}$\\
		$\mathrm{var}$ $arr = iter.\mathrm{toArray}$\\
		$\mathrm{val}$ $n = 6$\\
			\For {$j = 0 \to (arr.\mathrm{length}-n)$}
			{
				$words ::= arr.\mathrm{slice}( j, j+n).\mathrm{mkString}(''\ '')$\\
			}
		$words.\mathrm{iterator}$\\
	}
\end{algorithm}

\subsubsection{Multiple-length n-grams}
The multiple-length n-gram method has been proven to be effective to increase the detection accuracy in the area of system call-based HIDS \cite{wespi2000intrusion}\cite{marceau2001characterizing}\cite{khreich2017anomaly}. For instance, Creech et al. used multiple sliding windows with various sizes to scan the complete system call trace; the generated multiple-length n-grams were used for feature extraction, and notable detection accuracy was achieved\cite{creech2014semantic}. Therefore, we have implemented the multiple-length n-gram method with Spark for feature extraction.

The process of feature extraction with the multiple-length n-gram method is described as follows. Let $\bbbh$ = \{$\mathcal{T}_1$,  $\mathcal{T}_2$, ..., $\mathcal{T}_m$\} be the complete dataset of $m$ system call traces $\mathcal{T}$; and $\mathcal{T}$ = \{$S_1$, $S_2$, ..., $S_k$\} be a trace of $k$ system calls $S$. Then for a trace $\mathcal{T}$, a set of sliding windows ranging from size 1 to $n$ are applied to traverse from the beginning to the end of trace $\mathcal{T}$ to extract contiguous system call sequences. In this case, the maximum window size $n$ allowed is $k$, which is the length of trace $\mathcal{T}$. The contiguous system call sequences extracted at this stage form words $W$ of length 1 to $n$. Therefore, if $n$ is equal to 6, for trace $\mathcal{T}$ the generated document $\mathcal{D}$ that contains the extracted words $W$ of system calls is,

\noindent$\mathcal{D}$ = \{$S_1$, $S_2$, $S_3$, ..., $S_k$, \\
$S_1$$S_2$, $S_2$$S_3$, ..., $S_{k-1}$$S_k$, \\
$S_1$$\to$$S_3$, $S_2$$\to$$S_4$, ..., $S_{k-2}$$S_{k-1}$$S_k$, \\
$S_1$$\to$$S_4$, $S_2$$\to$$S_5$, ..., $S_{k-3}$$S_{k-2}$$S_{k-1}$$S_k$, \\
$S_1$$\to$$S_5$, $S_2$$\to$$S_6$, ..., $S_{k-4}$$S_{k-3}$$S_{k-2}$$S_{k-1}$$S_k$, \\
$S_1$$\to$$S_6$, $S_2$$\to$$S_7$, ..., $S_{k-5}$$S_{k-4}$$S_{k-3}$$S_{k-2}$$S_{k-1}$$S_k$\}.

\noindent Algorithm \ref{WordsGenMultiple} implements the multiple-length n-gram method with the sliding window algorithm with multiple window sizes. In Algorithm \ref{WordsGenMultiple}, the maximum window size $n$ is set to 6, thus the length of the extracted words can range from 1 to 6. When the maximum window size $n$ is set to one, the algorithm is the same with the single-length n-gram method.

\begin{algorithm}
	\caption{Generate words of system calls with the multiple-length n-gram method from one RDD partition}
	\label{WordsGenMultiple}
	\DontPrintSemicolon
	\SetAlgoLined
	\SetKwProg{Def}{def}{$=$}{end}
	\Def{$\mathrm{WordsGen[T:ClassTag]}(iter:\mathrm{Iterator[T]})$}
	{
		$\mathrm{var}\ words = \mathrm{List[String]()}$\\
		$\mathrm{var}$ $arr = iter.\mathrm{toArray}$\\
		$\mathrm{val}$ $n = 6$\\
		\For {$i = 0 \to (n-1)$}
		{
			\For {$j = 0 \to (arr.\mathrm{length}-i-1)$}
			{
				$words ::= arr.\mathrm{slice}( j, j+i+1).\mathrm{mkString}$\\
			}
		}
		$words.\mathrm{iterator}$\\
	}
\end{algorithm}

\subsubsection{Constructing feature vectors with TF-IDF}
``Term frequency-inverse document frequency (TF-IDF)'' is a feature extraction scheme commonly used in text-based information retrieval to measure the importance of a word in a document and the relevant corpus \cite{leskovec2014mining}. The term frequency (TF) is the number of times that the word appears in the document; the document frequency (DF) is the number of documents that contain the word \cite{hristidis2010ranked}. Using only TF may over-emphasize frequent words, as a frequent word in the corpus may carry little special information about a particular document; the inverse document frequency (IDF) can measure the quantity of information that a word provides about a particular document \cite{Sparkguide}. Thus the TF-IDF value of a word increases according to its frequency in the document and is adjusted by the frequency in the corpus. We have implemented TF-IDF with Spark to construct feature vectors. The process is described as follows. Let $\bbbd$ = \{$\mathcal{D}_1$,  $\mathcal{D}_2$, ..., $\mathcal{D}_n$\} denotes the complete corpus of $n$ documents $\mathcal{D}$. Term frequency $TF(W, \mathcal{D})$ is the number of times that a word $W$ appears in document $\mathcal{D}$, while document frequency $DF(W, \bbbd)$ is the number of documents that contain the word $W$ \cite{danesh2007improve}. In this case, the IDF is defined as,
\begin{equation}\label{eq1}
IDF(W, \bbbd) = \log \frac{n + 1}{DF(W, \bbbd) + 1}
\end{equation}
The TF-IDF is defined as,
\begin{equation}\label{eq2}
TFIDF(W, \mathcal{D}, \bbbd) = TF(W, \mathcal{D}) \cdot IDF(W, \bbbd)
\end{equation}

\begin{algorithm}[!b]
	\caption{Count the number of distinct words for an RDD of vectors of words}
	\label{alg5}
	\KwIn{An RDD of vectors of words $vects$}
	\KwOut{The number of distinct words $wordsNum$}
	\DontPrintSemicolon
	\SetAlgoLined
	\SetKwProg{Def}{def}{$=$}{end}
	\Def{$\mathrm{NumOfDistinctWords}(vects:\mathrm{ RDD[Seq[String]]):Int}$}
	{
		$\mathrm{val}$ $wordsNum =$ $ vects.\mathrm{flatMap\{identity\}.map}(x=>(x,1)).\mathrm{reduceByKey({\_}+{\_}).count.toInt}$\\
		\Return{$wordsNum$}\\
	}
\end{algorithm}
TF and IDF are implemented in \textit{HashingTF} and \textit{IDF} in Spark. The TF-IDF feature vectors generated from the training and testing system call traces are also treated as Spark RDDs. The \textit{HashingTF} utilizes feature hashing in machine learning \cite{weinberger2009feature}. In the proposed approach, to reduce the chance of hash collisions, the dimension number of the target feature vectors is set to the number of distinct words in the corpus of the training dataset. Algorithm \ref{alg5} is implemented to count the number of distinct words for an RDD of vectors of words. Algorithm \ref{alg3} is implemented to generate TF feature vectors from an RDD of vectors of words as sparse vectors. Algorithm \ref{alg4}\cite{Sparkguide} is implemented to generate TF-IDF feature vectors from an RDD of vectors of words as sparse vectors. 
In practice, the real-time system call traces are assumed to be gathered by the IDS trace by trace. Thus the whole testing dataset cannot be processed at one-time, instead only one single trace can be handled each time. In this case, for the testing traces, only TF feature vectors can be generated from them, and the IDF model is unobtainable. Therefore, algorithm \ref{alg4} is applied to the training dataset only. To generate the TF-IDF feature vectors of the testing dataset, firstly the TF feature vectors are generated using algorithm \ref{alg3}, and then adjusted by the Spark IDF model generated from the training dataset.

\begin{algorithm}[!t]
	\caption{Generate TF feature vectors from an RDD of vectors of words as sparse vectors}
	\label{alg3}
	\KwIn{An RDD of vectors of words $vects$, the number of dimensions $num$}
	\KwOut{Sparse TF feature vectors $tf$}
	\DontPrintSemicolon
	\SetAlgoLined
	\SetKwProg{Def}{def}{$=$}{end}
	\Def{$\mathrm{TfGen}(vects: \mathrm{RDD[Seq[String]]}, num:\mathrm{Int}): \mathrm{RDD[org.apache.spark.mllib.linalg.Vector]}$}
	{
		$\mathrm{val}$ $hashingTF$ = $\mathrm{new}$ $\mathrm{HashingTF}(num)$\\
		$\mathrm{val}$ $tf = hashingTF.\mathrm{transform}(vects).\mathrm{cache()}$\\
		\Return{$tf$}\\
	}
\end{algorithm}
\begin{algorithm}[!t]
	\caption{Generate TF-IDF feature vectors from an RDD of vectors of words as sparse vectors}
	\label{alg4}
	\KwIn{An RDD of vectors of words $vects$, the number of dimensions $num$}
	\KwOut{Sparse TF-IDF feature vectors $tfidf$}
	\DontPrintSemicolon
	\SetAlgoLined
	\SetKwProg{Def}{def}{$=$}{end}
	\Def{$\mathrm{TfidfGen}(vects: \mathrm{RDD[Seq[String]]}, num:\mathrm{Int}): \mathrm{RDD[org.apache.spark.mllib.linalg.Vector]}$}
	{
		$\mathrm{val}$ $hashingTF =$ $\mathrm{new}$ $\mathrm{HashingTF}(num)$\\
		$\mathrm{val}$ $tf = hashingTF.\mathrm{transform}(vects).\mathrm{cache()}$\\
		$\mathrm{val}$ $idf = $ $\mathrm{new}$ $\mathrm{IDF().fit}(tf)$\\
		$\mathrm{val}$ $tfidf =$ $idf.\mathrm{transform}(tf)$\\
		\Return{$tfidf$}\\
	}
\end{algorithm}

\subsection{Classifier training and prediction}
The training and prediction processes of machine learning classifiers can be classified as the optimization problem,
\begin{equation}
\min_{\wv \in\R^d} \; f(\wv)
\end{equation}
where $\wv$ is the weight vector with $d$ items. The objective function $f(\wv)$ can be written as,
\begin{equation}
f(\wv) := \lambda\, R(\wv) +
\frac1n \sum_{i=1}^n L(\wv;\x_i,y_i)
\end{equation}
where $\x_i\in\R^d,1\le i\le n$ is the training data, $y_i\in\R$ is the prediction label, $L(\wv; \x, y)$ is the loss function, $R(\wv)$ is the regularization term, and $\lambda$ is the regularization term parameter. The loss function measures the deviation between the prediction and the actuality; the regularization term controls the model complexity and uses the parameter to reduce the deviation while avoiding over-fitting. $\boldsymbol\ell1$ and $\boldsymbol\ell2$ are two commonly used regularization terms,

\begin{equation}
\boldsymbol\ell1:R(\wv):= \|\wv\|_1 \,\,\,\,\,\, \boldsymbol\ell2:R(\wv):= \frac{1}{2}\|\wv\|_2^2
\end{equation}

We use logistic regression and linear support vector machine for training and prediction. For the logistic regression classifier, the LBFGS algorithm is used for training; for the linear support vector machine classifier, the stochastic gradient descent method (SGD) and the $\boldsymbol\ell2$ regularization method are used for training.

\subsubsection{LBFGS logistic regression classifier (LR-LBFGS)}
Logistic regression is a widely used linear classification method. The loss function of logistic regression is,
\begin{equation}
L(\wv;\x,y) :=  \log(1+\exp( -y \wv^T \x))
\end{equation}
For $\x$, the trained model uses the logistic function to predict,
\begin{equation}
f(\wv^T \x) = \frac{1}{1 + e^{-\wv^T \x}}
\end{equation}
If $f(\wv^T \x)$ is greater than a given threshold, then the prediction is positive, otherwise it is negative. We use the LBFGS algorithm and $\boldsymbol\ell2$ regularization for training of the logistic regression classifier. The LBFGS algorithm is an optimization algorithm based on the Quasi-Newton method \cite{8681070}. The LBFGS algorithm uses limited computer memory to approximates the Broyden-Fletcher-Goldfarb-Shanno (BFGS) algorithm. The LBFGS algorithm is claimed to have fast convergence speed. Therefore, we use the LBFGS algorithm for training of the logistic regression classifier.

\subsubsection{Linear support vector machine classifier (SVM)}
Linear Support Vector Machine (SVM) is a linear method that is commonly used for classification of large-scale data. The loss function of SVM is the hinge loss,
\begin{equation}
L(\wv;\x,y) := \max \{0, 1-y \wv^T \x \}, \quad y \in \{-1, +1\}
\end{equation}
The SVM is predicted based on the value of $\wv^T \x$. If $\wv^T \x \geq 0$, the prediction is positive. If $\wv^T \x < 0$, the prediction is negative. We use the stochastic gradient descent method (SGD) and the $\boldsymbol\ell2$ regularization method for the training of SVM classifier.

\subsubsection{The training and prediction processes}
\begin{enumerate}[label=(\arabic*),leftmargin=0em,itemindent=1.5em,labelwidth=\itemindent,labelsep=1mm]
    \item In the processes of classifier training and prediction, firstly the attack traces and normal traces of ADFA-LD are loaded from the Google cloud storage system. The \textit{distinct} transformation, which can produce a new RDD with only distinct elements, is called to eliminate repetitive traces for both of the attack traces and normal traces. For both of the attack traces and normal traces, 60\% of them are randomly selected as the training data, and the other 40\% of them are selected as the testing data. 
	
    \item To get the number of dimensions $num$ for functions \textit{TfGen} and \textit{TfidfGen}, the attack training traces and normal training traces are combined using the \textit{union} method in Spark, followed by applying the \textit{distinct} transformation to eliminate repetitive traces. Then the function \textit{VectsGen} is applied to the combined traces to generate vectors of words, followed by applying the function \textit{NumOfDistinctWords} to get the number of distinct words, i.e., the number of dimensions.
    \begin{algorithm}[!htb]
    	\caption{Classifier training and prediction with LR-LBFGS}
    	\label{LBFGS}
    	\KwIn{The dataset for training $trainingData$, the dataset for testing $testingData$}
    	\KwOut{The value of AUC $AUC$}
    	\DontPrintSemicolon
    	\SetAlgoLined
    	$\mathrm{val}$ $model =$ $\mathrm{new}$ $\mathrm{LogisticRegressionWithLBFGS().run}(trainingData)$\\
    	$model$$\mathrm{.clearThreshold()}$\\
    	$\mathrm{val}$ $predictionAndLabels = testingData.\mathrm{map}$ 
    	$\{ \mathrm{case}\ \mathrm{LabeledPoint}(label, features) =>$
    	$\mathrm{val}\ prediction = model.\mathrm{predict}(features)$
    	$(prediction, label)\}$
    	
    	$\mathrm{val}$ $metrics =$ $\mathrm{new}$ $\mathrm{BinaryClassificationMetrics}$$(predictionAndLabels)$\\
    	$\mathrm{val}$ $AUC = metrics.\mathrm{areaUnderROC()}$\\
    	\Return{$AUC$}
    \end{algorithm}
    \item For both of the attack training data and normal training data, the function \textit{VectsGen} is called to generate vectors of words, followed by applying the function \textit{TfidfGen} to generate TF-IDF feature vectors. The attack feature vectors are labeled point 1, and the normal feature vectors are labeled point 0. Then the labeled feature vectors are combined using the \textit{union} method in Spark to form $trainingData$, which is the dataset for training the classifiers.
    \item In practice, as the real-time system call traces are assumed to be gathered by the IDS trace by trace, thus the whole testing dataset cannot be processed at one-time, instead only one single trace can be processed each time. Therefore, for both of the attack testing data and normal testing data, the function \textit{VectsGen} is called to generate vectors of words, followed by applying the function \textit{TfGen} to generate TF feature vectors. Then the function \textit{TfGen} is applied to the vectors of words formed in step (2) to generate TF feature vectors from the training dataset, followed by generating the relevant Spark IDF model of the training dataset. Then this IDF model is used to adjust the attack testing TF feature vectors and normal testing TF feature vectors. The (adjusted) attack feature vectors are labeled point 1, and the (adjusted) normal feature vectors are labeled point 0. 
    The labeled feature vectors are combined using the \textit{union} method in Spark to form $testingData$, which is the dataset for testing the classifiers.
	\item Algorithm \ref{LBFGS} and algorithm \ref{SVM}\cite{Sparkguide} implement the training and prediction with the LR-LBFGS classifier and the SVM classifier, respectively. The $trainingData$ and $testingData$ are prepared for training and prediction. For the LR-LBFGS classifier, the \textit{LogisticRegressionWithLBFGS} method is used for training; for the SVM classifier, the \textit{SVMWithSGD} method is used for training and the number of iterations is set to 100. Finally, the predicted labels are compared with the labels of the testing feature vectors to evaluate the detection accuracy. For these two algorithms, the AUC value $AUC$ is returned using the \textit{BinaryClassificationMetrics} method and the \textit{areaUnderROC} method.
\end{enumerate}

\begin{algorithm}[!htb]
	\caption{Classifier training and prediction with linear SVM}
	\label{SVM}
	\KwIn{The dataset for training $trainingData$, the dataset for testing $testingData$}
	\KwOut{The value of AUC $AUC$}
	\DontPrintSemicolon
	\SetAlgoLined
	$\mathrm{val}\ model = \mathrm{SVMWithSGD.train}(trainingData, 100)$\\
	$model.\mathrm{clearThreshold()}$\\
	$\mathrm{val}\ scoreAndLabels = testingData.\mathrm{map}$\
	$\{ point =>
	\mathrm{val}\ score = model.\mathrm{predict}(point.\mathrm{features})
	(score, point.\mathrm{label})\}$\\
	$\mathrm{val}\ metrics = \mathrm{new}\  \mathrm{BinaryClassificationMetrics}(scoreAndLabels)$\\
	$\mathrm{val}\ AUC = metrics.\mathrm{areaUnderROC}()$\\
	\Return{$AUC$}
\end{algorithm}

\section{Experiments}
\subsection{The environment for experiments}
The experiments are launched with Dataproc on the Google Compute Engine \cite{krishnan2015google} cloud computing environment. Dataproc provides pre-installed and administrated Apache Hadoop and Spark services with a user-friendly API for configuration. Dataproc can automatically and quickly create and manage clusters of virtual machines. There are three modes to deploy a cluster, i.e., ``single node (one master and zero workers)", ``standard (one master and multiple workers)", or ``high availability (three masters and numerous workers)". Virtual machines of a cluster can connect with each other with internal IP networking. Users can customize the number of virtual CPUs, the memory capacity, the sizes and types of disks, and the region/zone where the cluster to be deployed. The created master node contains the HDFS NameNode and the YARN ResourceManager, and can be accessed with SSH; each of the worker nodes includes an HDFS DataNode and a YARN NodeManager. Spark jobs can be submitted via the Dataproc API, where the jobs' output can be accessed; the graphs of the network, disk, and CPU utilization can also be viewed. Via the API, the created cluster can be scaled up or down regarding the number of worker nodes. Therefore, a cluster can be scaled multiple times with simple operations, which is flexible to test the scalability of the proposed approach. 

In our experiments, a free account has been set up firstly. Standard clusters (one master, multiple workers) are created within the \textit{global} region and \textit{us-central1-c} zone. For a cluster, the master node has one vCPU with 3.75GB memory, and the primary disk size is 100GB; each of the worker nodes has the same machine type with the master node. Spark version 2.2.1 on Dataproc is used for the experiments, and the number of Spark executors can be flexibly extended from 1 to 6.

\subsection{The dataset for experiments}
As most of the datasets utilized for measuring system call-based HIDS were created in the last century and cannot represent modern attack approaches, Creech et al. compiled the new publicly available ADFA-LD dataset \cite{creech2014semantic}. In their method, Ubuntu Linux is the host operating system, which represents a contemporary Linux server and provides multiple functionalities with a few vulnerabilities. The dataset has three subsets of raw system call traces, including 833 traces of normal data, 4372 traces of validation data, and 746 traces of attack data. Training and validation traces were gathered during normal activities of the operating system. The attack traces were gathered under the cyber attack environment. Attack methods include ``brute force password guessing", ``add new superuser", ``Java Meterpreter payload", ``Linux Meterpreter payload", and ``C100 webshell" \cite{creech2013generation}. The ADFA-LD dataset was claimed to be more challenging to investigate than the traditional datasets since the modern Linux environment has become more complicated than before \cite{creech2013generation}. Thus, the ADFA-LD dataset is a new benchmark for analyzing system call-based HIDS and is utilized in our experiments. 

\subsection{The experiment processes}
In our experiments, the attack traces and normal traces of ADFA-LD are utilized. Multiple files in the dataset are combined into less number of files for experiments. The dataset is prepared and stored in a bucket of Google's cloud storage system. With the Google cloud storage, users can store and retrieve large-scale data flexibly, regarding the time and the location. The default storage class of the bucket used for experiments is Multi-Regional, and the location is ASIA. During an experiment, after logging into the master node with SSH and launching the Spark shell, the attack traces and normal traces of ADFA-LD are firstly retrieved from the bucket of Google's cloud storage system, followed by performing the complete training and testing phases for intrusion detection.
\begin{figure}[!htb]
	\centering
	\tcbox[boxrule=0.3mm,boxsep=0mm,colback=white]{\includegraphics[width=0.44\textwidth]{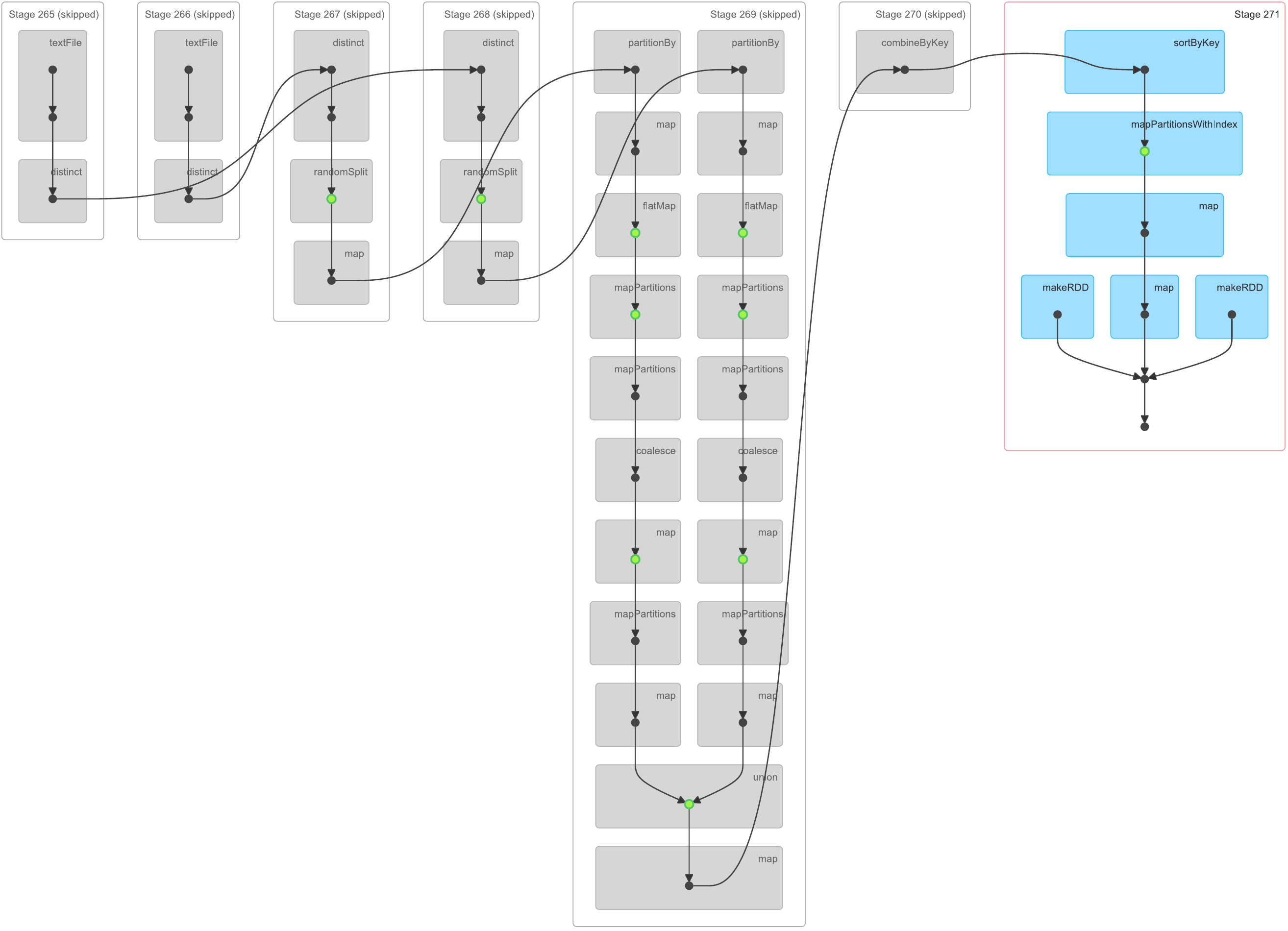}}
	\caption{The DAG formed using the RDD-based method with the LR-LBFGS classifier.}
	\label{SCADSDAG}
\end{figure}

\subsection{Experiment results}
Figure \ref{SCADSDAG} demonstrates the directed acyclic graph (DAG) formed using the multiple-length n-gram method and the LR-LBFGS classifier in an experiment. The DAG has 271 stages, which represent the complete experiment from the reading of data files to the returning of the AUC values. 

The effectiveness of the multiple-length n-gram method and the TF-IDF method are tested in the experiments. To test the effectiveness of using the multiple-length n-gram method for feature extraction, by using the LR-LBFGS classifier, the AUC values achieved using the multiple-length n-gram method are compared with the AUC values achieved using the single-length n-gram method. To test the effectiveness of using the TF-IDF method to construct feature vectors, for the testing traces, we have compared the AUC values achieved using only the TF feature vectors with using the TF-IDF feature vectors adjusted by the Spark IDF model generated from the training traces. The multiple-length n-gram method and the LR-LBFGS classifier are used in this case. 

The experiments also compare the LR-LBFGS classifier with the SVM classifier in terms of the detection accuracy and the scalability. The criteria of comparison are the AUC values and the average (executor) task time achieved when using these two kinds of classifiers.

\subsubsection{Evaluation regarding the detection accuracy}
The detection accuracy of a system call-based HIDS can be evaluated with the True Positive Rate (TPR) and False Positive Rate (FPR) criteria, which are defined as \cite{abed2015applying},
\begin{equation}
\mathrm{TPR=\frac{TP}{TP + FN}}\hspace{1em}
\mathrm{FPR=\frac{FP}{FP + TN}}
\end{equation}

\begin{enumerate}[label=(\arabic*),leftmargin=0em,itemindent=1.5em,labelwidth=\itemindent,labelsep=1mm]
	\item
	True Positive (TP): The label of a system call trace is attack and the prediction is also attack. The attack trace is detected correctly.
	\item
	False Positive (FP): The label of a trace is normal but the prediction is attack. A normal system call trace is falsely detected as an attack trace.
	\item
	True Negative (TN): The label of a trace is normal and the prediction is also normal. The testing system call trace is correctly predicted as normal.
	\item
	False Negative (FN): The label of a sequence is attack but the prediction is normal. The system omits an attack system call trace.
\end{enumerate}

The Receiver Operating Characteristic (ROC) curve can provide an intuitive view of the relation between the false positive rate (FPR) and the true positive rate (TPR) \cite{sacca2014improved}. In the proposed approach, the performance regarding detection accuracy is evaluated using values of the area under the ROC curve (AUC). AUC can be a simple metric to provide an overall evaluation of the detection accuracy. Using AUC values, we have evaluated the performance regarding three aspects, i.e., the effectiveness of using the multiple-length n-gram method for feature extraction, the effectiveness of using the TF-IDF method to construct feature vectors, and the comparison of using the LR-LBFGS classifier with using the SVM classifier in terms of the detection accuracy.
\setlength\tabcolsep{2pt}

\begin{enumerate}[label=(\arabic*),leftmargin=0em,itemindent=1.5em,labelwidth=\itemindent,labelsep=1mm]
    \item To test the effectiveness of using the multiple-length n-gram method for feature extraction, under the circumstance of using the LR-LBFGS classifier, the AUC values achieved using the multiple-length n-gram method are compared with the AUC values achieved using the single-length n-gram method. For the single-length n-gram method, we have tested the performance with the length of n from 1 to 10. For each length of n, the AUC values of ten experiments are recorded and averaged in table \ref{AUCvaluesSingleN}. For the multiple-length n-gram method, we have tested the performance with the maximum length of n from 1 to 10. For each of the maximum lengths of n, the AUC values of ten experiments are recorded and averaged in table \ref{AUCvaluesMultipleN}. The averaged AUC values for both of the single-length n-gram method and the multiple-length n-gram method are depicted in figure \ref{AUCsChap4Original} for comparison. 
    \begin{table}[!t]
    	\scriptsize
    	\centering
    	\caption{The AUC values obtained from 100 experiments using the single-length n-gram method with the LR-LBFGS classifier.}
    	\label{AUCvaluesSingleN}
    	\begin{tabular}{|>{\centering}p{0.75cm}|c|c|c|c|c|c|c|c|c|c|c|}
    		\hline 
    		n & E.1 & E.2 & E.3 & E.4 & E.5 & E.6 & E.7 & E.8 & E.9 & E.10 & Avg.\tabularnewline
    		\hline \hline
    		1 & 0.937 & 0.939 & 0.908 & 0.954 & 0.945 & 0.896 & 0.939 & 0.944 & 0.958 & 0.943 & 0.936\tabularnewline
    		\hline 
    		2 & 0.979 & 0.968 & 0.956 & 0.952 & 0.964 & 0.982 & 0.975 & 0.970 & 0.970 & 0.967 & 0.968\tabularnewline
    		\hline 
    		3 & 0.978 & 0.982 & 0.968 & 0.975 & 0.979 & 0.972 & 0.977 & 0.981 & 0.974 & 0.987 & 0.977\tabularnewline
    		\hline 
    		4 & 0.978 & 0.977 & 0.982 & 0.970 & 0.976 & 0.965 & 0.979 & 0.976 & 0.981 & 0.979 & 0.976\tabularnewline
    		\hline 
    		5 & 0.979 & 0.978 & 0.986 & 0.977 & 0.983 & 0.975 & 0.975 & 0.987 & 0.986 & 0.983 & 0.981\tabularnewline
    		\hline 
    		6 & 0.981 & 0.968 & 0.947 & 0.978 & 0.983 & 0.985 & 0.978 & 0.977 & 0.978 & 0.983 & 0.976\tabularnewline
    		\hline 
    		7 & 0.987 & 0.976 & 0.977 & 0.979 & 0.978 & 0.974 & 0.975 & 0.976 & 0.972 & 0.975 & 0.977\tabularnewline
    		\hline 
    		8 & 0.956 & 0.965 & 0.955 & 0.966 & 0.972 & 0.971 & 0.981 & 0.988 & 0.966 & 0.979 & 0.970\tabularnewline
    		\hline 
    		9 & 0.972 & 0.967 & 0.963 & 0.967 & 0.969 & 0.943 & 0.958 & 0.952 & 0.970 & 0.968 & 0.963\tabularnewline
    		\hline 
    		10 & 0.936 & 0.958 & 0.964 & 0.962 & 0.947 & 0.965 & 0.954 & 0.959 & 0.966 & 0.960 & 0.957\tabularnewline
    		\hline 
    	\end{tabular}
    \end{table}
    \begin{table}[!t]
    	\scriptsize
    	\centering
    	\caption{The AUC values obtained from 100 experiments using the multiple-length n-gram method with the LR-LBFGS classifier.}
    	\label{AUCvaluesMultipleN}
    	\begin{tabular}{|c|c|c|c|c|c|c|c|c|c|c|c|}
    		\hline 
    		Max. n & E.1 & E.2 & E.3 & E.4 & E.5 & E.6 & E.7 & E.8 & E.9 & E.10 & Avg.\tabularnewline
    		\hline \hline
    		1 & 0.885 & 0.940 & 0.936 & 0.945 & 0.956 & 0.945 & 0.951 & 0.922 & 0.935 & 0.943 & 0.936\tabularnewline
    		\hline 
    		2 & 0.978 & 0.957 & 0.966 & 0.963 & 0.976 & 0.981 & 0.965 & 0.976 & 0.973 & 0.974 & 0.971\tabularnewline
    		\hline 
    		3 & 0.970 & 0.986 & 0.978 & 0.974 & 0.986 & 0.967 & 0.985 & 0.968 & 0.973 & 0.978 & 0.977\tabularnewline
    		\hline 
    		4 & 0.977 & 0.979 & 0.986 & 0.969 & 0.973 & 0.975 & 0.988 & 0.979 & 0.985 & 0.957 & 0.977\tabularnewline
    		\hline 
    		5 & 0.988 & 0.984 & 0.989 & 0.987 & 0.982 & 0.980 & 0.990 & 0.978 & 0.985 & 0.981 &  0.984\tabularnewline
    		\hline 
    		6 & 0.981 & 0.992 & 0.987 & 0.980 & 0.982 & 0.987 & 0.990 & 0.985 & 0.979 & 0.983 & 0.985\tabularnewline
    		\hline 
    		7 & 0.986 & 0.985 & 0.985 & 0.987 & 0.981 & 0.985 & 0.992 & 0.984 & 0.982 & 0.980 & 0.985\tabularnewline
    		\hline 
    		8 & 0.981 & 0.977 & 0.989 & 0.988 & 0.983 & 0.989 & 0.986 & 0.981 & 0.989 & 0.981 & 0.984\tabularnewline
    		\hline 
    		9 & 0.980 & 0.981 & 0.980 & 0.987 & 0.977 & 0.985 & 0.986 & 0.979 & 0.991 & 0.983 & 0.983\tabularnewline
    		\hline 
    		10 & 0.980 & 0.979 & 0.987 & 0.989 & 0.979 & 0.993 & 0.980 & 0.992 & 0.983 & 0.983 & 0.985\tabularnewline
    		\hline 
    	\end{tabular}
    \end{table}
    \begin{table}[!t]
    	\scriptsize
    	\centering
    	\caption{The AUC values obtained from 100 experiments using the multiple-length n-gram method with the LR-LBFGS classifier (only TF feature vectors are used for testing).}
    	\label{AUCvaluesTFonly}
    	\begin{tabular}{|c|c|c|c|c|c|c|c|c|c|c|c|}
    		\hline
    		Max. n & E.1 & E.2 & E.3 & E.4 & E.5 & E.6 & E.7 & E.8 & E.9 & E.10 & Avg.\tabularnewline
    		\hline \hline
    		1 & 0.807 & 0.841 & 0.818 & 0.739 & 0.756 & 0.935 & 0.770 & 0.658 & 0.813 & 0.817 & 0.795\tabularnewline
    		\hline 
    		2 & 0.836 & 0.878 & 0.859 & 0.890 & 0.929 & 0.886 & 0.844 & 0.906 & 0.850 & 0.860 & 0.874\tabularnewline
    		\hline 
    		3 & 0.928 & 0.926 & 0.938 & 0.947 & 0.973 & 0.948 & 0.957 & 0.925 & 0.938 & 0.935 & 0.942\tabularnewline
    		\hline 
    		4 & 0.913 & 0.973 & 0.940 & 0.904 & 0.974 & 0.913 & 0.957 & 0.964 & 0.946 & 0.944 & 0.943\tabularnewline
    		\hline 
    		5 & 0..912 & 0.971 & 0.874 & 0.895 & 0.979 & 0.970 & 0.970 & 0.912 & 0.981 & 0.928 & 0.939\tabularnewline
    		\hline 
    		6 & 0.968 & 0925 & 0.941 & 0.902 & 0.955 & 0.953 & 0.907 & 0.938 & 0.961 & 0.980 & 0.943\tabularnewline
    		\hline 
    		7 & 0.901 & 0.957 & 0.881 & 0.920 & 0.972 & 0.936 & 0.984 & 0.916 & 0.943 & 0.920 & 0.933\tabularnewline
    		\hline 
    		8 & 0.975 & 0.964 & 0.983 & 0.966 & 0.976 & 0.907 & 0.968 & 0.978 & 0.982 & 0.975 & 0.967\tabularnewline
    		\hline 
    		9 & 0.978 & 0.974 & 0.951 & 0.954 & 0.942 & 0.944 & 0.982 & 0.962 & 0.981 & 0.978 & 0.965\tabularnewline
    		\hline 
    		10 & 0.959 & 0.970 & 0.973 & 0.959 & 0.958 & 0.970 & 0.916 & 0.966 & 0.958 & 0.981 & 0.961\tabularnewline
    		\hline 
    	\end{tabular}
    \end{table}
    \begin{table}[!t]
    	\scriptsize
    	\centering
    	\caption{The AUC values obtained using the multiple-length n-gram method with the SVM classifier.}
    	\label{SVMtable}
    	\begin{tabular}{|c|c|c|c|c|c|c|c|c|c|c|c|}
    		\hline 
    		Max. n & E.1 & E.2 & E.3 & E.4 & E.5 & E.6 & E.7 & E.8 & E.9 & E.10 & Avg.\tabularnewline
    		\hline 
    		1 & 0.834 & 0.837 & 0.884 & 0.882 & 0.899 & 0.812 & 0.912 & 0.920 & 0.899 & 0.775 & 0.865\tabularnewline
    		\hline 
    		2 & 0.929 & 0.943 & 0.951 & 0.921 & 0.933 & 0.951 & 0.953 & 0.950 & 0.927 & 0.938 & 0.940\tabularnewline
    		\hline 
    		3 & 0.950 & 0.960 & 0.969 & 0.967 & 0.946 & 0.959 & 0.957 & 0.962 & 0.962 & 0.962 & 0.959\tabularnewline
    		\hline 
    		4 & 0.954 & 0.960 & 0.952 & 0.960 & 0.969 & 0.956 & 0.968 & 0.954 & 0.957 & 0.970 & 0.960\tabularnewline
    		\hline 
    		5 & 0.956 & 0.951 & 0.948 & 0.948 & 0.957 & 0.962 & 0.951 & 0.957 & 0.957 & 0.964 & 0.955\tabularnewline
    		\hline 
    		6 & 0.962 & 0.955 & 0.962 & 0.951 & 0.959 & 0.954 & 0.956 & 0.971 & 0.953 & 0.960 &  0.958\tabularnewline
    		\hline 
    		7 & 0.968 & 0.957 & 0.978 & 0.955 & 0.950 & 0.968 & 0.960 & 0.957 & 0.958 & 0.961 & 0.961\tabularnewline
    		\hline 
    		8 & 0.949 & 0.954 & 0.972 & 0.955 & 0.962 & 0.950 & 0.947 & 0.965 & 0.963 & 0.963 & 0.958\tabularnewline
    		\hline 
    		9 & 0.955 & 0.957 & 0.969 & 0.952 & 0.957 & 0.960 & 0.959 & 0.949 & 0.950 & 0.958 & 0.957\tabularnewline
    		\hline 
    		10 & 0.941 & 0.960 & 0.963 & 0.956 & 0.946 & 0.960 & 0.956 & 0.950 & 0.977 & 0.951 & 0.956\tabularnewline
    		\hline
    	\end{tabular}
    \end{table}
    Based on figure \ref{AUCsChap4Original}, for the multiple-length n-gram method, the AUC values reach a plateau when the maximum length of n is greater than 5. For the single-length n-gram method, the AUC values reach the maximum when n is equal to 5. Based on our implementation methods and experiment results, the multiple-length n-gram method outperforms the single-length n-gram method using the LR-LBFGS classifier when the sizes of n (or maximum n) range from 1 to 10.

\begin{figure}[!t]
	\centering
	\subfigure[AUC values obtained using the single-length n-gram method or the multiple-length n-gram method]{
		\includegraphics[width=0.4\textwidth]{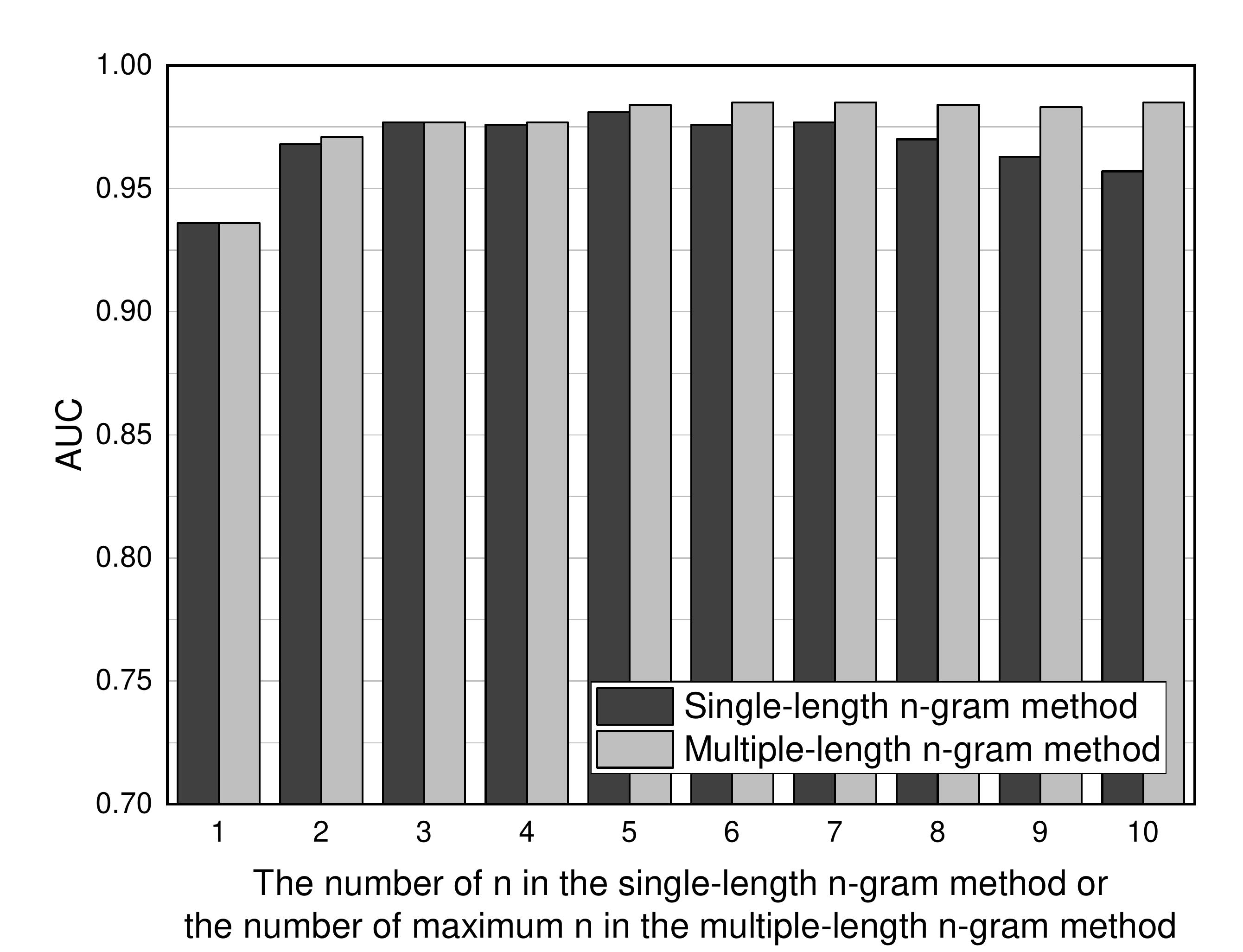}
		\label{AUCsChap4Original}
	}
	\subfigure[AUC values obtained using TF feature vectors only or TF-IDF feature vectors for testing traces]{
		\includegraphics[width=0.4\textwidth]{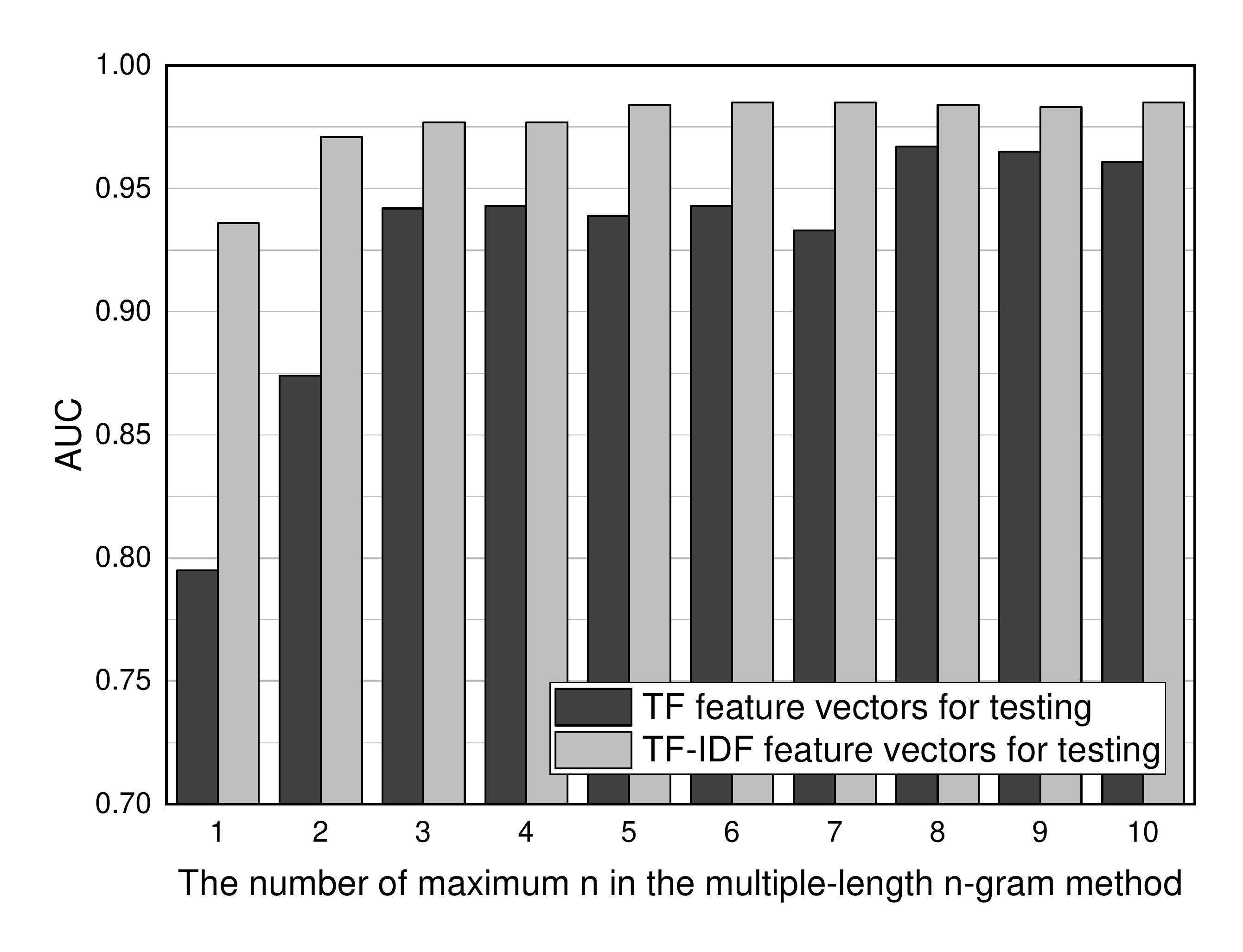}
		\label{AUCsChap4TFonly}
	}
	\caption{AUC values obtained in the experiments with the LR-LBFGS classifier.}
	\label{AUCsChap4}
\end{figure} 
\begin{figure}[!t]
	\centering
	\includegraphics[width=0.4\textwidth]{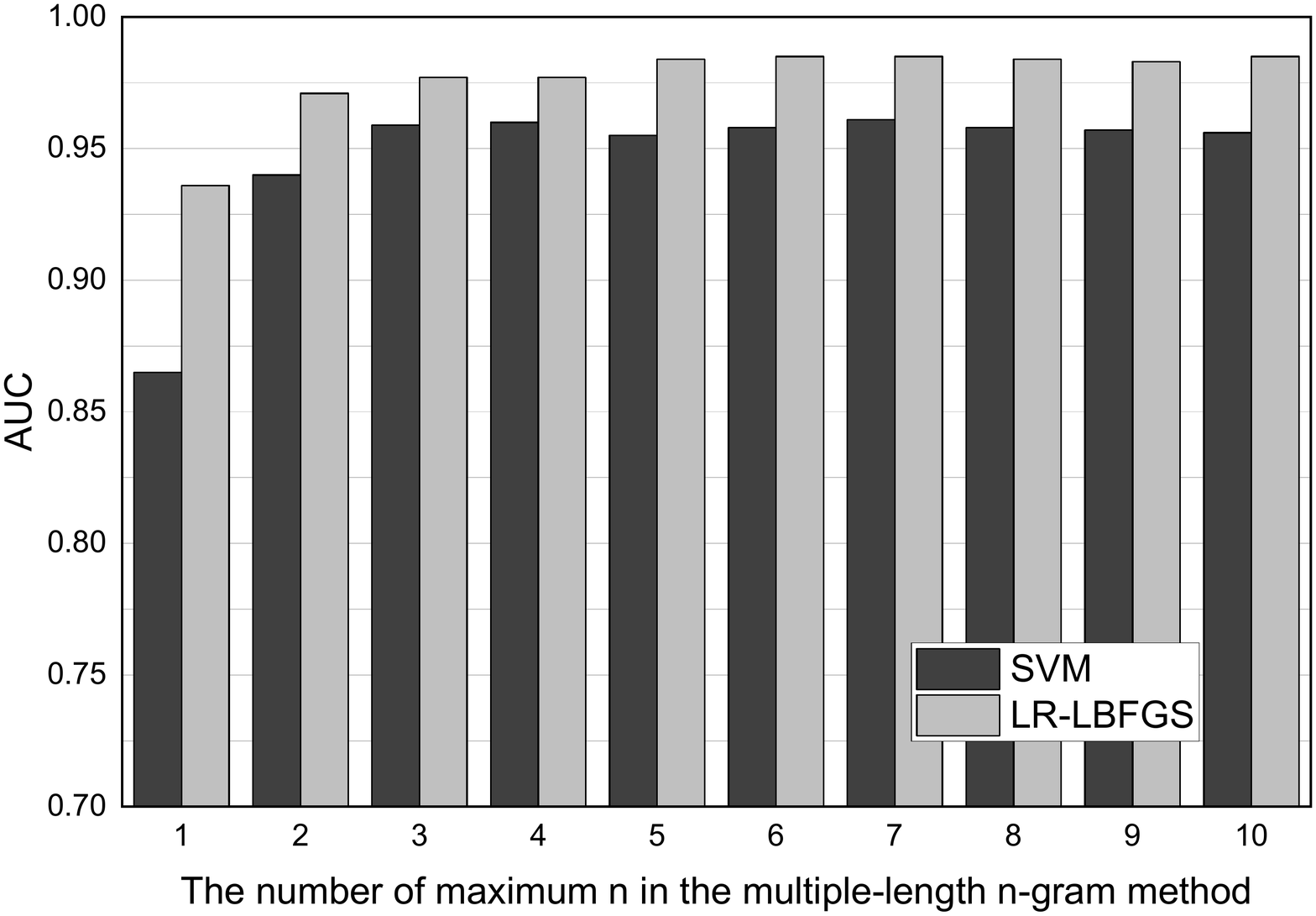}
	\caption{AUC values obtained with the multiple-length n-gram method and the SVM classifier.}
	\label{AUCSVMChap4}
\end{figure}
    \item To test the effectiveness of using the TF-IDF method to construct feature vectors, for the testing traces, we have compared the AUC values obtained using only the TF feature vectors with using the TF-IDF feature vectors adjusted by the Spark IDF model generated from the training traces. The multiple-length n-gram method and the LR-LBFGS classifier are used. The AUC values of using only the TF feature vectors for testing traces are recorded in table \ref{AUCvaluesTFonly}. For each of the maximum lengths of n, the AUC values of ten experiments are recorded and averaged. The averaged AUC values of these two cases (TF only and TF-IDF) are depicted in figure \ref{AUCsChap4TFonly}. Based on our implementation methods and experiment results, using only the TF feature vectors for testing shows unstable performance, and using Spark IDF model generated from the training traces to adjust the TF feature vectors of the testing traces is effective to improve the detection accuracy.
    
    \item The experiments compare the LR-LBFGS classifier and the SVM classifier regarding the AUC values. According to table \ref{SVMtable} and figure \ref{AUCSVMChap4}, for the LR-LBFGS classifier, the AUC values reach a plateau when the maximum length of n is greater than 5; for the SVM classifier, the AUC values reach a plateau when the maximum length of n is greater than 3. During this group of experiments, when the iteration number of SVM is set to 100, the LR-LBFGS classifier outperforms the SVM classifier regarding the detection accuracy.
\end{enumerate}

\begin{figure}[!htb]
	\centering
	\subfigure[LR-LBFGS classifier]{
		\includegraphics[width=0.4\textwidth]{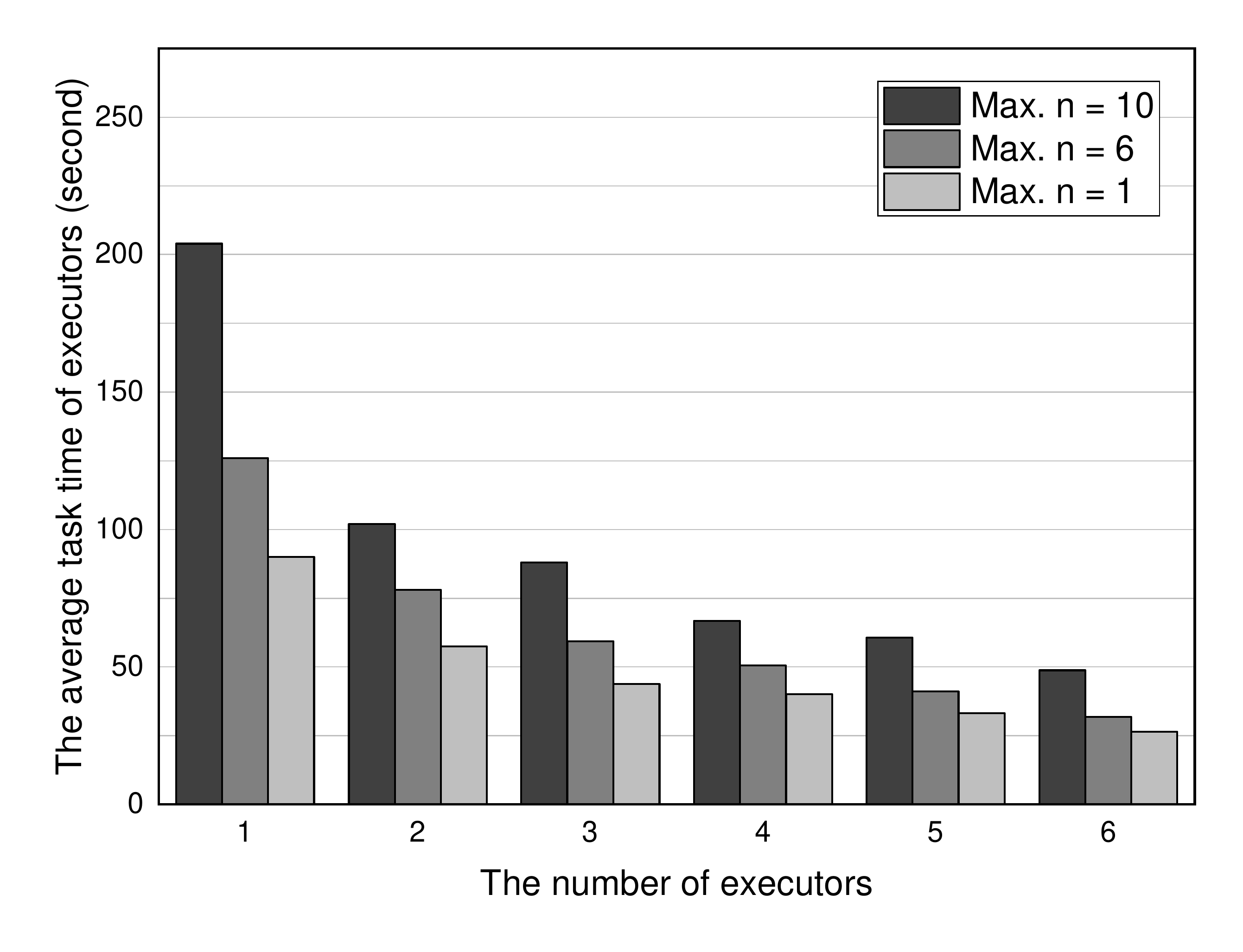}
		\label{ScalabilityChap4lbfgs}
	}
	\subfigure[SVM classifier]{
		\centering
		\includegraphics[width=0.4\textwidth]{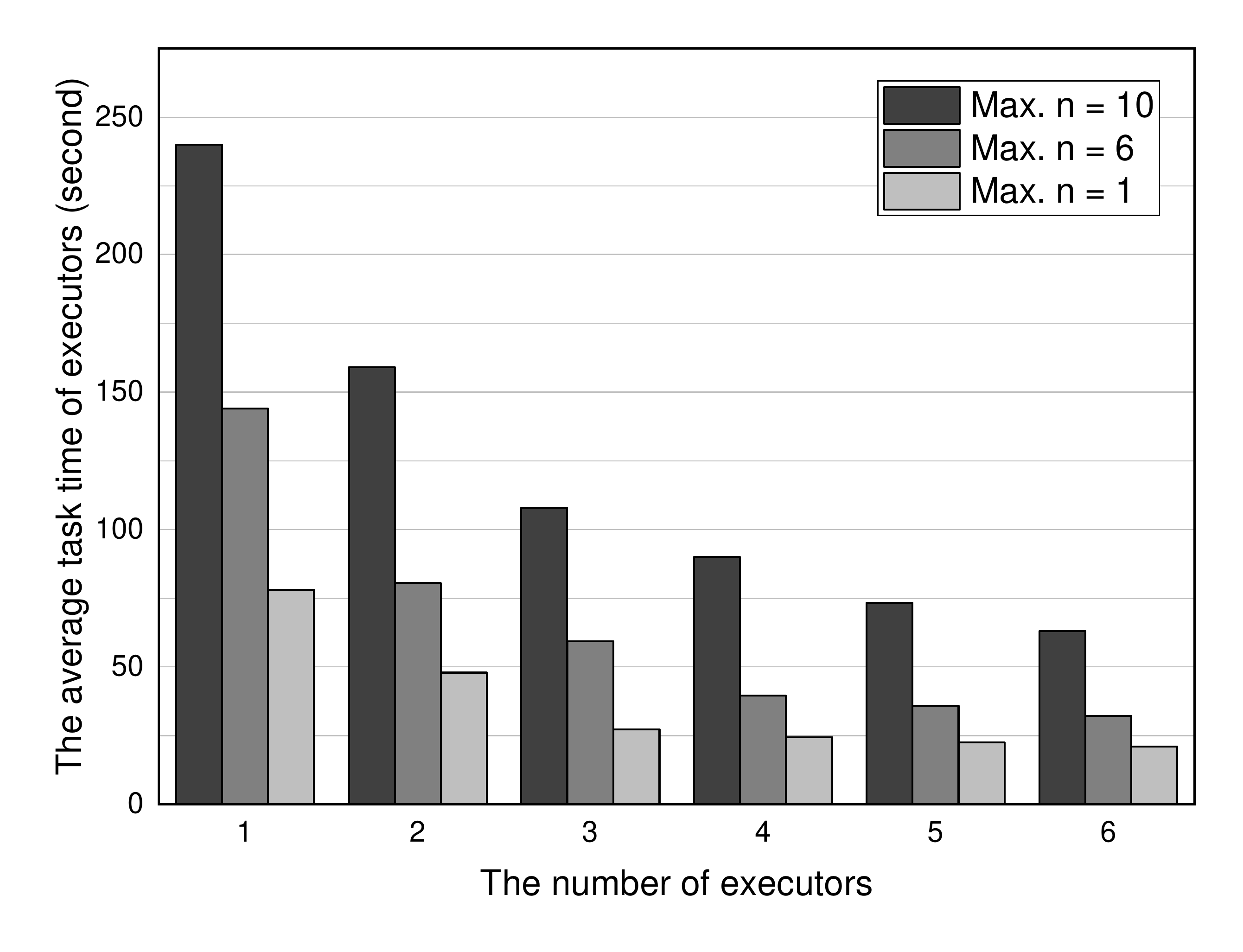}
		\label{ScalabilityChap4SVM}
	}
	\caption{The scalability of SCADS (the number of executors ranges from 1 to 6).}
	\label{Scalability4}
\end{figure} 
\subsubsection{Evaluation of the scalability}
Via the Dataproc API, the created Spark cluster can be flexibly scaled up or down to test the scalability of the proposed approach. As we use a free account of the Google Compute Engine, the number of Spark executors in our experiments ranges from 1 to 6. The experiments for the evaluation of scalability are divided into two groups. The first group uses the LR-LBFGS classifier, the second group uses the SVM classifier. The multiple-length n-gram method is used in both of the two groups, and the maximum length of n is equal to 1, 6, and 10, respectively. For both of the two groups of experiments, in each case of the number of Spark executors (1 to 6), we have executed the system three times, applying the multiple-length n-gram method with different maximum lengths of n (1, 6, and 10). The relevant average task time of Spark executors is recorded each time. Thus there are 54 times of experiment in each group of experiments. The results are averaged in figure \ref{Scalability4}. Based on figure \ref{Scalability4}, with the increase of the maximum length of n in the multiple-length n-gram method, the average task time of Spark executors increases as well. When the maximum length of n is equal to 10, the LR-LBFGS classifier is more efficient than the SVM classifier regarding the average task time of Spark executors. For both of the two classifiers, with the increase of the number of Spark executors, the average task time of Spark executors decreases apparently. Therefore, the approach of SCADS shows acceptable scalability.

\section{Conclusion and future works}
In this article, we have proposed SCADS, a scalable approach using Spark in the Google cloud for host-based intrusion detection system with system calls. Based on the experiment results, the proposed framework with Apache Spark and Google cloud computing services can improve the detection efficiency of traditional system call-based HIDS. Besides the algorithms used in our experiments, other existing intrusion detection algorithms may also be incorporated into this proposed framework with Apache Spark and Google cloud computing services. However, as the current experiments are conducted with a free account, the scale of computation is relatively limited. Therefore, in the future experiments, we will keep tunning the system to achieve better performance, and we will also consider upgrading the account to a paid one for the processing of real-time large-scale system call traces.

\ifCLASSOPTIONcaptionsoff
  \newpage
\fi
\bibliographystyle{IEEEtran}
\bibliography{mybibfile}

\end{document}